# Photonic realization of chiral hinge states in a Chern-insulator stack

Han-Rong Xia,[1] Jia-Zheng Li,[1] Si-Yu Yuan,[1] Meng Xiao,[1, 2*]

[1]Key Laboratory of Artificial Micro- and Nano-structures of Ministry of Education and School of Physics and Technology, Wuhan University, Wuhan 430072, China
[2]Wuhan Institute of Quantum Technology, Wuhan 430206, China

Corresponding Email [*]: phmxiao@whu.edu.cn

Higher-order topological insulators, as a novel family of topological phases, are a hot frontier in condensed matter physics due to their adherence to unconventional bulk-boundary correspondence. A three-dimensional second-order topological insulator can support one-dimensional modes along its hinges (dubbed as hinge states). Here, we present a simple and direct method to construct chiral hinge modes based on a Chern-insulator stack. We analyze the existence of the hinge modes through the nontrivial quadrupole indices, and then design a photonic crystal to realize the specific flowing pattern of the hinge mode in our model. The experimental results align well with full-wave simulations, clearly demonstrating the existence of chiral hinge states. We also verify the robustness of these hinge states against defects in our photonic system.

# I. INTRODUCTION

Over the past decades, the field of topological phases has experienced rapid development, emerging as a vibrant branch of condensed matter physics [1-3]. The traditional bulk-boundary correspondence (BBC), often taken as a defining property of topological phases, connects the bulk topological invariants in a $d$-dimensional system with the topologically protected $(d-1)$-dimensional boundary states, dubbed as the first-order topology [4-6]. Recently, a new class of topological phases, known as higher-order topological insulators (HOTIs), has attracted wide research interest [7-9]. Distinguished from the conventional topological insulators (TIs) obeying the first-order BBC, HOTIs can support lower-dimensional boundary states, such as one-dimensional (1D) hinge states in a second-order three-dimensional (3D) HOTI. The seminal work of Benalcazar et al. in 2017 formally introduced the concept of HOTIs [7] and triggered a research boom. The famous quantized quadrupole TI model (i.e. 2D BBH model) features a square lattice with a $\pi$ flux per plaquette and dimerized couplings along both the *x*- and *y*- directions. This model generalizes the concept of dipole moment to quadrupole moment in TIs and reveals in-gap corner states protected by a bulk topological quadrupole invariant [7]. Since the BBH model was proposed, there have been successive other approaches to realizing HOTIs such as a model with glide symmetry [10], 2D and 3D SSH models with quantized bulk polarization [11-13], etc. [14-19] Need to mention that there are also some extrinsic HOTIs based on boundary topological transitions [19] rather than the bulk topological invariants. These extrinsic HOTIs provide intuitive and straightforward strategies for realizing HOTIs by designing the surface orientation or boundary terminations. A consistent and systematic way for characterizing the topology of the aforementioned HOTIs (both intrinsic and extrinsic) under arbitrary orientations was proposed very recently [20].

Since 2017, a variety of concrete models on HOTIs have been extensively implemented in physical systems, including electronics [21,22], photonics [12,23-25] and acoustics [10,13-16,26-28]. The limitation on finding suitable materials and achieving straightforward detection leads to the slow development of experimental research in electronic systems. In contrast, classical wave systems, such as photonic and acoustic crystals, serve as exciting and popular platforms for exploring HOTIs, mainly due to their macroscopic scale, flexible sample fabrication and capability of direct wave function measurement. There have already been quite a few nice works for demonstrating topological insulators and semimetals in acoustic [29-36] and photonic [37-42] systems. Meanwhile, hinge states have also been successfully demonstrated via acoustic cavity-waveguide systems, both in higher-order topological semimetals [43-46] and insulators [19,47-51]. However, compared with the acoustic counterpart, the photonic realizations of hinge states lag far behind [52-54]. The challenge lies largely in the fact that the coupling in electromagnetic waves is far more difficult to engineer than acoustic systems where one can simply control the cross-section of the waveguides between cavities. As an integral component of classical wave topology, research on photonic hinge states urgently needs to be advanced. In addition, chiral hinge states, as a special type of hinge states, offer the possibility of guiding light unidirectionally along an arbitrarily designed route in a 3D system. This unique property is crucial in topologically robust optical communication. However, there is currently no realization of chiral hinge states that are immune against any defects and scatterers regardless of the (pseudo-)spin degree of freedom.

As a simple and effective approach to constructing photonic HOTI, here in this work, we implement the layer stacking system. Layer stacking with interlayer coupling tuning is a common method to construct novel topological phases in both photonic and acoustic crystals, which has been demonstrated successfully in previous works [49,52,55-58]. We start with 2D Chern insulators, and by stacking 2D Chern insulators with different topological numbers, we confine the edge states of the 2D Chern insulators also in the stacking direction. As a consequence, the edge states become the hinge states that propagate unidirectionally along the edge of a 3D Chern insulator stack. In other words, we realize the chiral photonic hinge states of HOTIs. We offer a tight-binding model and show that the windings of the quadrupole moments characterize the nontrivial topology of this system. We have also implemented this scheme within a microwave photonic crystal, and provide convincing experimental demonstration. We should mention that although hinge states have been demonstrated in previous works [19,21,22,43-54], these hinge states are not chiral in the sense that they can be scattered to the back-propagating hinge states or into the bulk. To the best of our knowledge, our work realizes the chiral hinge states in classical waves (both photonic and acoustic systems) for the first time.

This paper is organized as follows. In Sec. II, we give the framework of our design. To be more specific, we first construct two Chern insulators with opposite Chern numbers, stack them alternately, and then tune the interlayer coupling to form a Su-Schrieffer-Heeger (SSH) like model in the vertical direction. In Sec. III, a theoretical analysis of the hinge states in our system is provided. The windings of the quadrupole moments determine the flowing pattern of the hinge states in our system. In Sec. IV, we design a photonic crystal to implement the model described in Sec. II and perform full-wave simulations to show the dispersion and expected excitation results. Experimental demonstration as well as the robustness against defects of hinge states are provided in Sec. V. Finally, we summarize in Sec. VI.

## II. DESIGN PRINCIPAL OF THE HINGE STATES

Our approach starts with 2D Chern insulators. As illustrated in Fig. 1(a), layer A has a Chern number $-C$ (upper panel) and layer B exhibits a Chern number $+C$ (lower panel). These two Chern insulators can be related to each other by time reversal operation. (In the experiments, this operation can be simply implemented by reversing the direction of the external magnetic field.) Nonzero Chern numbers lead to the presence of gapless chiral edge states flowing anticlockwise or clockwise along the boundaries [59]. When coupling two layers of 2D Chern insulators with opposite Chern numbers, the combined system becomes topologically trivial, and the chiral edge states interact and their dispersions become gapped. Stacking such a two-layer system along the $z$ direction [Fig. 1(b)] naturally ends in a trivial system where the Chern vector [60] is $(0,0,0)$. However, the situation becomes nontrivial when engineering the interlayer coupling as shown in Fig. 1(c). Similar to the SSH model, here we consider dimerized interlayer coupling as denoted by $t_1$ and $t_2$. The situation is greatly simplified in the limit of $t_1/t_2 \to 0$ and $t_1/t_2 \to \infty$. When $t_1/t_2 \to 0$, there is one dangling layer A that is decoupled from the bulk stack. Such a layer A has a nonzero Chern number and should support a chiral edge state propagating counterclockwise on the upper edge as denoted by the blue arrows in Fig. 1(c). Correspondingly, there is also another chiral edge state propagating clockwise on the lower surface of the stack as shown with the pink arrows in Fig. 1(c). These two chiral edge states propagate along the hinges of the 3D stack and are hence dubbed as chiral hinge states. Note here, these two chiral hinge states are localized at opposite surfaces of the stack and hence decoupled from each other for a thick enough stack. On the other hand, when $t_1/t_2 \to \infty$, there should be no chiral hinge states. One can naturally envision that a topological transition occurs from $t_1/t_2 \to 0$ to $t_1/t_2 \to \infty$. However, the emergence of the pair of chiral hinge states and the topological transition cannot be captured by the traditional topological invariant Chern vector since it vanishes in both cases.

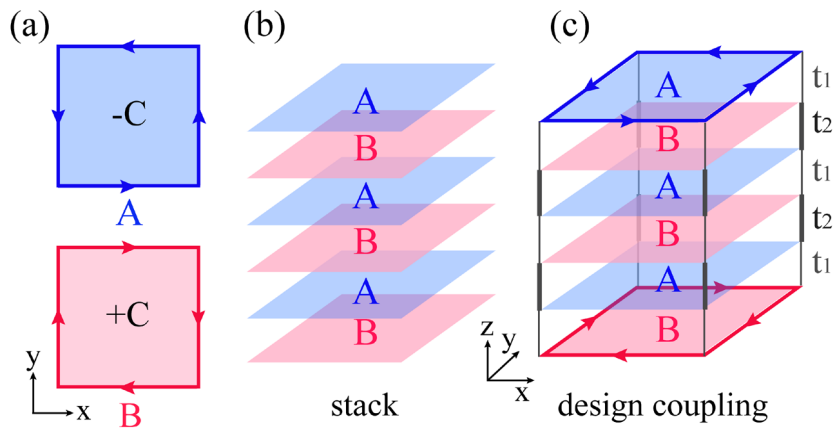

FIG. 1. Schematic illustration for the construction of hinge states. (a) Two kinds of monolayers with opposite Chern numbers, i.e., $-C$ for layer A (blue square) and $+C$ for layer B (pink square), supporting chiral edge states (denoted by solid arrowed lines) propagating anticlockwise and clockwise, respectively. (b) Stacking equal pieces of layers A and B alternatively along the $z$ direction, the Chern vector of the composite vanishes. (c) Illustration of considering dimerized interlayer coupling in the $z$ direction and truncating the system at weak interlayer coupling $t_1$. The chiral edge states propagate only at the boundaries of the outermost layers, thus forming chiral hinge states in a 3D stacking system.

### III. EFFECTIVE HAMILTONIAN APPROACH

The topological characterization of stacking systems has been discussed extensively for electronic systems [61-65]. Here to keep the text concise, we implement an effective Hamiltonian that respects all the symmetries possessed by our system to illustrate the nontrivial topology. The Hamiltonian describing our system, i.e., a Chern-insulator stack with $C_4$ and PT symmetry, can be written as,

$$H = \begin{pmatrix} h_{2d}(m, k_x, k_y) & h_c(k_z) \\ h_c^\dagger(k_z) & h_{2d}(-m, k_x, k_y) \end{pmatrix}, \quad (1)$$

where $h_{2d}(m, k_x, k_y)$ is the effective Hamiltonian of the monolayer A and $h_{2d}(-m, k_x, k_y)$ represents that of layer B, $h_c(k_z)$ describes the interlayer coupling. Here $m$ denotes the mass term which we intentionally invert such that layer A and layer B possess opposite Chern numbers. $h_{2d}(m, k_x, k_y)$ is constructed based on the rotation eigenvalues of our experimental sample as (Details are provided in the Supplementary Materials Sec. I),

$$h_{2d}(m, k_x, k_y) = \begin{pmatrix} -2 & 0 & H_{13} & H_{14} \\ 0 & 2 - \frac{5}{2}(\cos k_x + \cos k_y) & H_{23} & H_{24} \\ H_{13}^* & H_{23}^* & m & H_{34} \\ H_{14}^* & H_{24}^* & H_{34}^* & -m \end{pmatrix} \quad (2)$$

where $H_{13} = it_{1,3}\sin k_x + t_{1,3}\sin k_y$, $H_{14} = -it_{1,3}\sin k_x + t_{1,3}\sin k_y$, $H_{23} = -it_{2,3}\sin k_x + t_{2,3}\sin k_y$, $H_{24} = -it_{2,3}\sin k_x - t_{2,3}\sin k_y$ and $H_{34} = \frac{1}{2}\cos k_y - \frac{1}{2}\cos k_x$.

$$h_c = \frac{1}{2}(\sigma_0 - \sigma_3) \otimes \sigma_1(t_1 + t_2 e^{-ik_z}). \quad (3)$$

Here $m$, $t_{1,3}$, and $t_{2,3}$ are the controlling parameters such that the rotational eigenvalues of the effective Hamiltonian are in the same order as the sample by solving the Maxwell equations. Their corresponding values can be fitted via the band structure of the sample. Having constructed the effective Hamiltonian, we then proceed to calculate the topological invariants of this system. First, we calculate the Chern vector and as expected, it vanishes. In other words, there should be no surface of chiral boundary state for a finite sample. Since the Chern vector of our system vanishes, we consider checking other topological invariants. We find that the nontrivial topology should be characterized by the winding of quadrupole moments instead [66,67]. The quadrupole moments along the three principal directions for the nontrivial case $t_1 < t_2$ are shown in Fig. 2(a-c). Here the windings of the quadrupole moments along the $x$, $y$ and $z$ directions are $\Delta_x = 1$, $\Delta_y = -1$ and $\Delta_z = 0$, respectively. Such windings of quadrupole moments, i.e. quadrupole indices, indicate that the chiral hinge modes will emerge on the face in the $z$ direction with a double-loop pattern shown in Fig. 2(d). In addition, we have also calculated the quadrupole indices for the trivial case with $t_1 > t_2$. As expected, $\Delta_x = \Delta_y = \Delta_z = 0$. (The results are provided in the Supplementary Materials Sec. II.)

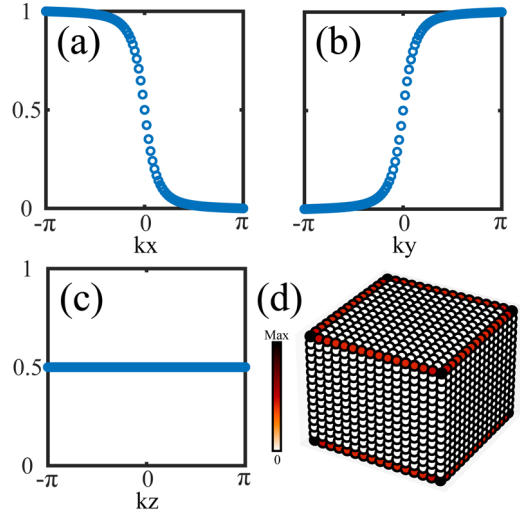

FIG. 2. Effective model. (a-c) The windings of quadrupole moments along three principal directions characterize the pattern of chiral hinge modes. Here, $t_1 = 0.2$, $t_2 = 0.5$, $m = 1$, $t_{1,3} = 1 + i$ and $t_{2,3} = 2 + 2i$. (d) Amplitude of an eigenstate pattern shows the emergence of hinge states on the $xy$-plane surface.

## IV. PHOTONIC REALIZATION

Gyromagnetic photonic crystals (GPCs) are superb platforms for implementing our idea, owing to their natural response to external magnetic fields that enables Chern insulators. We design a square-lattice GPC system whose unit cell is schematically illustrated in Fig. 3(a). The gyromagnetic cylinders inside each layer (the blue and pink cylinders) are magnetized along the positive or negative $z$ direction as denoted by the white arrows. Except for the magnetization direction, the other parameters of the gyromagnetic cylinders inside all the layers are the same. The magnetization of the gyromagnetic cylinders is realized by sandwiching them between two pieces of magnets (marked in gray) with properly chosen orientations. We intentionally shift the unit cell along the *x-y* direction for half the unit cell to minimize the interaction among magnets. The magnets are embedded inside the perfect electric conductors (PEC, the yellow plates) so as to fix their positions. The interlayer interaction is introduced by the holes drilled in the PEC plates with the radius $r_1$ or $r_2$. The height $h$ of the PEC plates together with the radius of the holes control the strength of interlayer coupling.

We carry out full-wave simulations via the commercial finite-element software COMSOL MULTIPHYSIC$^{TM}$. The magnets' surfaces are covered by copper foil, so electric fields that penetrate the magnets are negligible. In the simulations, we filled the regions of magnets with PEC in order to reduce the computation load. The responses of cylinders A and B to the external magnetic field are captured by the relative permeability tensors [68,69],

$$\overleftrightarrow{\mu}_A = \begin{pmatrix} \mu_r & i\kappa & 0 \\ -i\kappa & \mu_r & 0 \\ 0 & 0 & 1 \end{pmatrix}, \overleftrightarrow{\mu}_B = \begin{pmatrix} \mu_r & -i\kappa & 0 \\ i\kappa & \mu_r & 0 \\ 0 & 0 & 1 \end{pmatrix}, \qquad (4)$$

where $\mu_r = 1 + \frac{(\omega_0 + i\alpha\omega)\omega_m}{(\omega_0 + i\alpha\omega)^2 - \omega^2}$, $\kappa = \frac{\omega\omega_m}{(\omega_0 + i\alpha\omega)^2 - \omega^2}$, $\omega_m = \gamma\mu_0 M_s$, $\omega_0 = \gamma\mu_0 H_z$, $\gamma$ is the gyromagnetic ratio, $\alpha$ is the damping coefficient, $\omega$ is the operating frequency, $4\pi M_s = 1400\text{G}$ is the saturation magnetization, and $\mu_0 H_z = 0.06\text{T}$ is the external magnetic flux density along the $z$ direction used in our experiments. We start with the band structure for a single layer A, where no holes are drilled on the PEC plates. Figure S1(c) shows the band structure for this quasi-2D system with the Chern number marked on each band. The yellow shadow highlights the band gap of interest. The total Chern number of the bands below this gap submits to $-1$ which hence verifies that a single layer A is a Chern insulator. As layer B is the time reversal counterpart of layer A, its band structure should be the same as layer A while the Chern number for the band gap of interest is $+1$.

We proceed to show the band structure for a finite stacking system. We set the radii of the smaller holes $r_1 = 0.8\text{mm}$, while the bigger ones $r_2 = 4.5\text{mm}$ to construct a significant interlayer coupling strength difference. We then stack three copies of these pairs to form the system as illustrated in the side view in Fig. 3(b). To calculate the dispersion of hinge states in such a six-layer system, we consider a supercell that is periodic in $y$ direction and finite in $x$ direction, as depicted in the top view of Fig. 3(c). Note here the blue and pink cylinders still belong to different layers in the 3D sample as the information of layers is indistinguishable from the top view. There are $N_x$ unit cells in layer A while $N_x - 1$ in layer B. The distance from the outermost cylinders to the boundary is denoted as $x_0$, which determines the detailed dispersion of the chiral edge state for each isolated layer. The numerical band structure (for $N_x = 17$) is plotted in Fig. 3(d). Here the gray, black and green dots represent the bulk, surface and hinge states, respectively. The pink strip highlights the frequency region where there are only the hinge states. The field distribution of the eigenstate marked by the yellow star is shown in the Supplementary Materials Sec. III, from which we can clearly see that the modes are well confined at the boundary of the outermost layers, i.e., hinge of the stacking.

The localization of fields and dispersion of the hinge modes can also be extracted from the field distributions under a point source excitation. To match the setup in the experiments, we now turn to a sample that is also finite along the *y* direction. The number of unit cells along the *y* direction of layer A is also set as $N_y = 17$ and the sample exhibits $C_4$ symmetry about the *z* direction. Since the four boundaries are equivalent for the system, we only extract the field data from the edges along the *y* direction. On each layer, we place a dipole source to excite the waves near the boundary of that layer and then scan the field. (Detailed setup can be found in the Supplementary Materials Sec. IV.) After measuring the field distribution inside each layer, we perform a fast Fourier transform (FFT) to obtain the band dispersion as shown in Fig. 3(e). The bright color indicates the corresponding well-excited mode. It is clear that most of the excited waves propagate backward in the first layer (L1, top layer), third layer (L3) and fifth layer (L5), while the waves in the other layers are predominantly forward propagating. This asymmetric propagating feature is inherited from the chiral edge states of the 2D Chern insulators. The interlayer coupling introduces a mixture between chiral edge states that propagate in opposite directions, thus a small amount of excited wave can also propagate forward (backward) in L3 and L5 (L2 and L4). In addition, the interlayer coupling also opens a gap between excited surface waves as indicated by the two green dashed lines in L2-L5. This gap is absent for L1 and L6 where the gapless chiral hinge states are present. Since the coupling between the L1 and L2 (L6 and L5) is pretty weak (the radii of the smaller holes are only $r_1 = 0.8$ mm), the leakage of the hinge state that localized on L1 (L6) to L2 (L5) is pretty small. Note here the FFT spectra of the hinge states in L1 and L6 agree well with the band structure for solving the eigensystem of a monolayer (the solid green lines). That also explains why the measurements performed on L2 (L5) can barely see the signal of the hinge states. Similarly, the measurements performed on L1 and L6 can barely see the back-propagating waves even for frequencies above and below the surface band gap (indicated by the dashed lines).

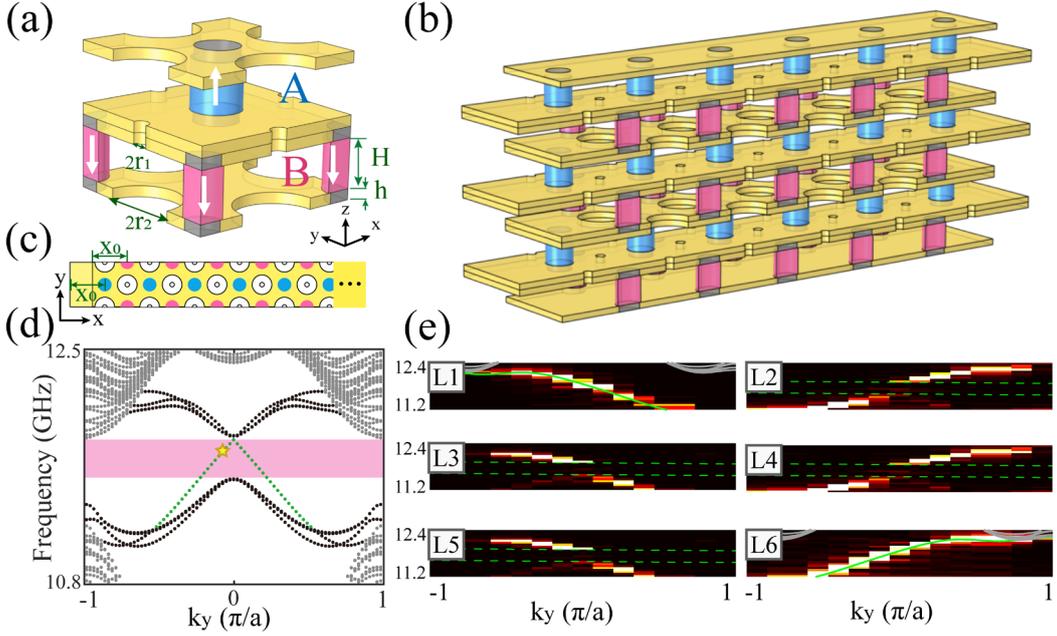

FIG. 3. Photonic crystal design and full-wave simulations. (a) Schematic diagram for the unit cell of a stacked-Chern-insulator crystal. Here the blue and pink cylinders are the gyromagnetic dielectric with the white arrows denoting the magnetization direction, the gray pillars are the magnets and the yellow plates (each with a height of $h = 1$mm) are PEC. Holes with radii $r_1 = 0.8$mm or $r_2 = 4.5$mm are drilled on the PEC plates. The crystal has a square lattice in the $x$-$y$ plane and the lattice constant is $a = 16$mm. The gyromagnetic cylinders have relative permittivity $\varepsilon_r = 13.5$, height $H = 4.5$mm, and diameter $D = 4.5$mm. (b, c) Side view and top view of the sample which is periodic along the $y$ direction, finite along the $x$ direction and has 6 layers along the $z$ direction. Here $N_x = 17$ and for simplicity, we do not show all the lattices in the $x$ direction. The distance between the outermost GPCs and the boundary is set as $x_0 = 11.7$mm. (d) Band structure of the system in (b, c). The frequency range of bulk and boundary gap is marked by the pink-shaded region where the eigenfields are well confined at the hinges of the stack. (e) FFT spectra of the field measured numerically on all the layers. The green solid lines denote the monolayer project band, while the dashed green lines outline the boundary gap region obtained in (d). The details of the setup are provided in the Supplementary Materials Sec. IV.

## V. EXPERIMENTAL DEMONSTRATION IN MICROWAVES

To experimentally verify our proposal, we fabricate a six-layer sample, as shown in Fig. 4(a). Each layer is a Chern insulator consisting of yttrium iron garnet (YIG) ferrite rods magnetized by small permanent magnets with diameter $d = 4$mm right above and below the YIG. A close-up view of one layer is photographed in Fig. 4(b), with the top cover layer glided several lattices away from its edges for better illustration. The magnets are embedded in the copper-clad plate to fix their positions. To form controllable edge states in every layer, we place four metal strips outside the YIG rods and adjust their positions one by one, ensuring the dispersions of edge states in all layers are approximately the same. The measured band dispersions of the edge states at boundary $Y_1$ [as denoted in Fig. 4(b)] for all six uncoupled monolayers are plotted in the Supplementary Materials Sec. V. From there, we can see that the edge states band dispersions are almost identical for the layers L1, L3 and L5 (L2, L4 and L6).

Same as Fig. 3(d), we then place the source at boundary $Y_1$ and probe the electric field distribution layer by layer. The FFT results of fields extracted from each $Y_1$ are shown in Fig. 4(c) which match well with the simulated results in Fig. 3(d). There are obvious signatures of the band gap in layers L2-L5 while the dispersions of the hinge states in layers L1 and L6 are gapless. As a supplement, the existence of the hinge states can also be directly revealed by the field distribution. In Fig. 4(d), we present the electric field distributions at frequency $f_0 = 11.8$GHz, which is inside the gap of both the boundary and bulk states. Since the bulk is also gapped, here we only show the field distribution inside the outmost unit cells. For uncoupled layers, the excited edge states can propagate for around 10 unit cells before eventually attenuating due to intrinsic loss [see Fig. S5]. In contrast, for stacking systems, the edge states decay quickly inside layers L2-L5 while still propagate inside layers L1 and L6. Hence, the field distributions in Fig. 4(d) also verify the emergence of hinge states at the outmost layers, i.e., L1 and L6.

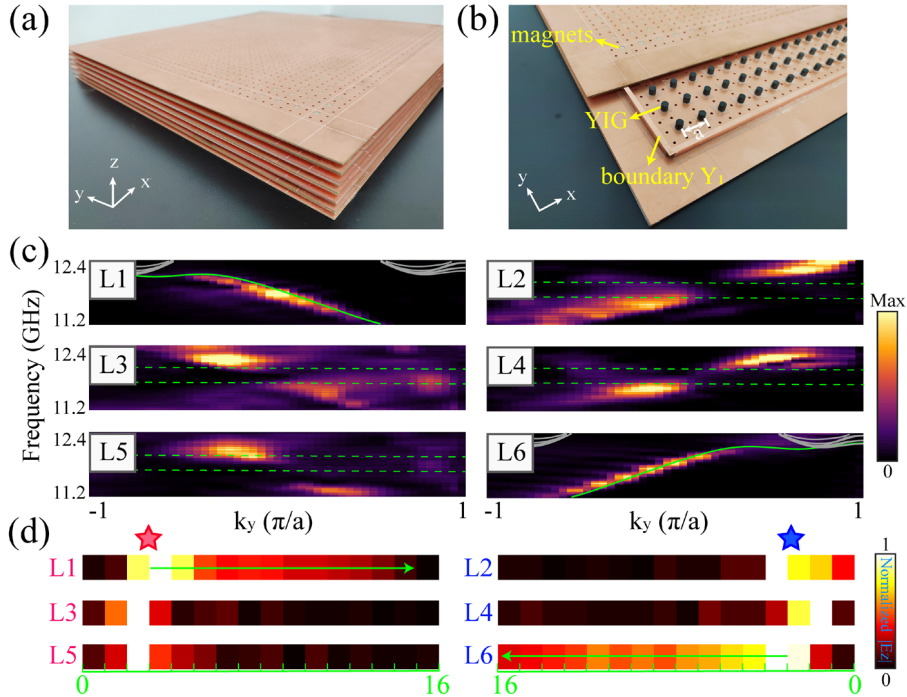

FIG. 4. Experimental demonstration of the chiral hinge states. (a, b) Photography of our experimental sample. The whole 3D sample is shown in (a), and a close look for one layer is given in (b) with the cover layer glided a certain distance. (c) FFT spectra of measured electric field for excitation and field scanning layer by layer. (d) The normalized field distributions of boundary $Y_1$ for each layer at frequency $f_0 =$ 11.8GHz. The colored stars mark the locations of the source. The green grids below highlight the number of the unit cell. The detailed geometric parameters are the same as in Fig. 3.

Due to the topological protection, the chiral hinge states are robust against defects and sharp corners. If the losses are small enough, the chiral hinge states can circle the whole boundaries of the outermost layers. To verify the robustness of these chiral hinge states, we place a metallic obstacle at layer $L6$, as illustrated in Fig. 5(a) with the green dashed rectangle. Here the blue star marks the position of the source. The measured field distribution is provided in Fig. 5(b). It clearly shows that the hinge states can go around the obstacle and propagate ahead, verifying the topological protection against defects. The decaying of the hinge state is owing to the intrinsic losses of our setup. In Fig. S6, we also show that the hinge states in the top layer ($L6$) can propagate around the corner counterclockwise.

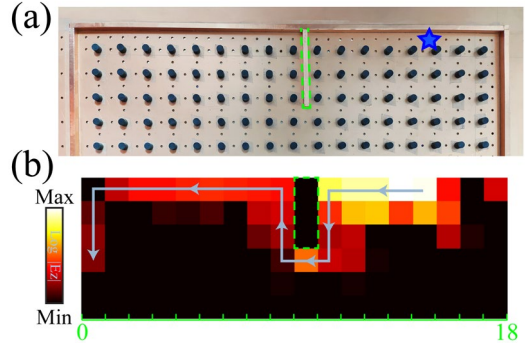

FIG. 5. The robustness of the hinge states against a metallic defect. (a) Photography of the top layer with a metallic obstacle (highlighted with the green dashed rectangle) which is about three periods long. (b) The corresponding field distribution. The blue star in (a) mark the place of source in the experiment. The working frequency is at $f_0 =$ 11.8GHz, and all the other parameters are the same as those in Fig. 4.

## VI. SUMMARY

To conclude, we construct chiral hinge states based on a Chern-insulator stack with dimerized interlayer couplings. The bulk of our system is trivial in the sense that its Chern vector vanishes. However, when the stacking is truncated at the weak interlayer coupling, our system can exhibit chiral hinge states that propagate unidirectionally on the top and bottom layers. We show that such chiral hinge states originate from the nontrivial windings of the quadrupole moments. We measured the dispersion and analyzed the field pattern of these chiral hinge states. Our results show good consistency between experimental measurements and full-wave simulations. In addition, we also demonstrate that the chiral hinge modes are robust against scattering due to the topological protection. The chiral hinge modes in our system can lead to potential applications that need to unidirectionally guide waves along a designed route in a 3D space.

**ACKNOWLEDGEMENT**

This work is supported by the National Key Research and Development Program of China (Grant No. 2022YFA1404900), the National Natural Science Foundation of China (Grant No. 12334015, Grant No. 12274332 and Grant No.12321161645).